\documentclass[12pt]{article}

\usepackage[dvips]{graphicx}

\usepackage{pdproc}

\usepackage{amsmath,amsbsy,amssymb,latexsym}

  \makeatletter 
  \def\@cite#1{[#1]} 
  \makeatother    
\newcommand\CL{{\mathcal L}}
\newcommand\CW{{\mathcal W}}
\newcommand\CF{{\mathcal F}}
\newcommand\CD{{\mathcal D}}
\newcommand\fD{{\mathfrak D}}

\newcommand\CN{{\mathcal N}}

\newcommand\CV{{\mathcal V}}
\newcommand\CG{{\mathcal G}}

\def\tr{\mathop{\rm tr}}
\def\Tr{\mathop{\rm Tr}}
\def\Im{\mathop{\rm Im}}
\def\Re{\mathop{\rm Re}}

\def\parity{\mathfrak R}

\begin{document}
\renewcommand{\thefootnote}{\alph{footnote}}
\begin{flushright}
OCU-PHYS-219\\
hep-th/0410132
\end{flushright}
\vspace{10mm}
\title{
$U(N)$
 Gauge Model and Partial Breaking of $\CN=2$ Supersymmetry
\footnote{Talk given by H.I.
at 
the 12th International Conference on Supersymmetry
and Unification of Fundamental Interactions (SUSY2004), 
June 17-23, 2004, Tsukuba, Japan.}
}
\vspace{5mm}
\author{ K. FUJIWARA~,~~ H. ITOYAMA
}
\address{ 
Department of Mathematics and Physics,
Graduate School of Science,
Osaka City University\\
3-3-138, Sugimoto, Sumiyoshi-ku, Osaka, 558-8585, Japan
\\ {\rm E-mail: fujiwara@sci.osaka-cu.ac.jp,~
itoyama@sci.osaka-cu.ac.jp}
}

\author{M. SAKAGUCHI} 
\address{Osaka City University Advanced Mathematical Institute
(OCAMI)\\ 
3-3-138, Sugimoto, Sumiyoshi-ku, Osaka, 558-8585, Japan
\\ {\rm E-mail: msakaguc@sci.osaka-cu.ac.jp}
}

\abstract{
We briefly review a construction
of $\CN=2$ supersymmetric $U(N)$ gauge model
in which rigid ${\cal N}=2$
 supersymmetry is spontaneously broken to ${\cal N}=1$.
This model
generalizes the abelian model considered by Antoniadis,
Patouche and Taylor.
We discuss the conditions on
the vacua of the model with
partial supersymmetry breaking.
}

\normalsize\baselineskip=15pt

\section{Introduction}

Let us recall that
 partial breaking of extended rigid supersymmetries
 appears not possible on the basis of the positivity of
  the  supersymmetry charge algebra:
\begin{eqnarray}
\left\{ \bar{Q}_{\alpha}^i,Q_{j\dot{\alpha}} \right\}=2(\boldsymbol{1})_{\alpha \dot{\alpha}}\delta_{~j}^{i} H.
\end{eqnarray}
In fact, if $\left. Q_1|0 \right>=0$, one concludes
 $\left. H|0 \right>=0$ and  $\left. Q_i|0 \right> =0$ for all $i$.
If $\left.Q_1|0 \right> \neq 0$,
 then $\left. H|0 \right> =\left. E|0\right>$ with $E>0$
 and $\left. Q_i|0 \right> \neq 0$ for all $i$.
 The loophole to this argument is  that
the use of the local version of the charge algebra is more appropriate
 in spontaneously broken symmetries and the most general supercurrent
 algebra  is 
\begin{eqnarray}
\left\{ \bar{Q}_{\dot{\alpha}}^j,\mathcal{S}_{\alpha i}^m (x) \right\}=2(\sigma^n)_{\alpha \dot{\alpha}} \delta_{~i}^{j} T_n^m(x)+(\sigma^m)_{\alpha \dot{\alpha}} C_i^j, \label{intro}
\end{eqnarray}
where $ \mathcal{S}_{\alpha i}^m $ and  $ T_n^m $ are
 the supercurrents and the energy momentum tensor respectively.
 We have denoted by $C_i^j$
a field independent constant matrix  permitted
 by  the constraints.
This last term does not modify the supersymmetry algebra acting on the fields.
The abelian model of \cite{APT} and 
our nonabelian generalization \cite{FIS1}
provide a concrete example of this local algebra
within linear realization from the point of view of the action principle.

\section{$\CN=2$ $U(N)$ Gauge Model}
We consider a $U(N)$ gauge model which is composed of
a set of $\CN=1$ chiral superfields $\Phi=\Phi^at_a$
and that of $\CN=1$ vector superfield strengths $\CW=\CW^at_a$
in the adjoint representation of $U(N)$.
The $t_a$ form $u(N)$ algebra $[t_a, t_b]=f^c_{ab}t_c$.
The kinetic term for $\Phi$ is specified by 
the K\"ahler potential 
\begin{eqnarray}
K(A^a, A^{*a})=\frac{i}{2} (A^a \CF^*_a
- A^{*a} \CF_a),~~~\CF_a\equiv \partial_a\CF,
\label{Kahler potential}
\end{eqnarray}
where $\CF$ is an analytic function of $A$,
and thus the K\"ahler metric takes the form $g_{ab^*}=\Im \CF_{ab}$.
The gauging of the $U(N)$ isometry is accomplished
by following the method \cite{gauging}
and using the
Killing potential
\begin{eqnarray}
\fD_a=
-\frac{1}{2}(\CF_bf^b_{ac} A^{*c}+\CF_{b}^*f^b_{ac}A^c) .
\label{Killing potential}
\end{eqnarray}
For 
gauge fields
we introduce the non-canonical kinetic action:
$-\frac{i}{4}\int d^2\theta \tau_{ab}\CW^a\CW^b + c.c.$,
where $\tau_{ab}$ is an analytic function of $\Phi$.
In addition, we include the superpotential $W(\Phi)$
and the Fayet-Iliopoulos D-term $\sqrt{2}\xi D^0$.
The index $0$ refers to the overall $u(1)$.

We found that our action 
is
invariant
under the $\parity$ action
\begin{eqnarray}
\left(
  \begin{array}{c}
    \lambda^a   \\
    \psi^a   \\
  \end{array}
\right)
\to
\left(
  \begin{array}{c}
    \psi^a  \\
    -\lambda^a   \\
  \end{array}
\right)~&,&~~~
D^c+\frac{1}{2}g^{cd}\fD_d
\to
-( D^c+\frac{1}{2}g^{cd}\fD_d
)  ~,
\nonumber\\
\xi\to -\xi~&,&~~~
F^a+g^{ac^*}\partial_{c^*} W^*
\to
F^{*b}+g^{db^*}\partial_{d}W       ~,
\label{R action}
\end{eqnarray}
if we choose $\tau_{ab}$ and $W(\Phi)$ as
\begin{eqnarray}
\tau_{ab}=\CF_{ab},~~~W=eA^0+m\CF_0,
\end{eqnarray}
where $e$ and $m$ are real constants.
The $\parity$ action is a discrete element 
of the $SU(2)$ $R$-symmetry 
that acts as  an automorphism
of $\CN=2$ supersymmetry.
Because our action is invariant under
the $\parity$ action as well as
the $\CN=1$ supersymmetry
transformation $\delta_{\eta_1}$,
it is invariant under the second supersymmetry
$\delta_{\eta_2}=\parity\delta_{\eta_1}\parity^{-1}$.
Thus, our action is invariant under the $\CN=2$ supersymmetry.

As a result,
the action of our $\CN=2$ $U(N)$ gauge model
is
\begin{eqnarray}
\CL&=&
-g_{ab^*}\CD_mA^a\CD^mA^{*b}
-\frac{1}{4}
g_{ab}v_{mn}^av^{bmn}
-\frac{1}{8}
{\rm Re}
(\CF_{ab})\epsilon^{mnpq}v_{mn}^av_{pq}^b
\label{action}
\\&&
-\frac{1}{2}\CF_{ab}\lambda^a\sigma^m\CD_m\bar\lambda^b
-\frac{1}{2}\CF_{ab}^*\CD_m\lambda^a\sigma^m\bar\lambda^b
-\frac{1}{2}\CF_{ab}\psi^a\sigma^m\CD_m\bar\psi^b
-\frac{1}{2}\CF_{ab}^*\CD_m\psi^a\sigma^m\bar\psi^b
\nonumber\\&&
+g_{ab^*}F^aF^{*b}
+F^a\partial_a W
+F^{*a}\partial_{a^*}  W^*
+\frac{1}{2}
g_{ab}D^aD^b
+\frac{1}{2}D^a \left( \fD_a + 2\sqrt{2} \xi \delta_{a}^0 \right)
\nonumber\\&&
+(\frac{i}{4}\CF_{abc}F^{*c}-\frac{1}{2}\partial_a\partial_b W)
 \psi^a\psi^b
+\frac{i}{4}\CF_{abc}
F^c\lambda^a\lambda^b 
+\frac{1}{\sqrt{2}}(g_{ac^*}k_{b}^*{}^{c}+\frac{1}{2}\CF_{abc}D^c)
\psi^a\lambda^b
\nonumber\\&&
+(-\frac{i}{4}\CF^*_{abc}F^{c}-\frac{1}{2}\partial_{a^*}
\partial_{b^*}  W^*)
\bar\psi^a\bar\psi^b 
-\frac{i}{4}\CF^*_{abc}
F^{*c}\bar\lambda^a\bar\lambda^b
+\frac{1}{\sqrt{2}}(g_{ca^*}k_{b}{}^{c}+\frac{1}{2}\CF^*_{abc}D^c)
\bar\psi^a\bar\lambda^b
\nonumber\\&&
-i\frac{\sqrt{2}}{8}(
\CF_{abc}
\psi^c\sigma^n\bar\sigma^m\lambda^a
-\CF^*_{abc}
\bar\lambda^a\bar\sigma^m\sigma^n\bar\psi^c
)v_{mn}^b
-\frac{i}{8}\CF_{abcd}
\psi^c\psi^d\lambda^a\lambda^b
+\frac{i}{8}\CF^*_{abcd}
\bar\psi^c\bar\psi^d\bar\lambda^a\bar\lambda^b, 
\nonumber
\end{eqnarray}
where $\CD_m$ represents
the gauge covariant derivative.

\section{Extended Supersymmetry Transformation}
Combining the $\CN=1$ supersymmetry transformation
with the $\parity$ transformation,
we can construct the $\CN=2$ supersymmetry transformation
acting on our model:
\begin{eqnarray}
\boldsymbol{\delta} A^a &=&
\sqrt{2} \boldsymbol{\eta}_j \boldsymbol{\lambda}^{ja},
~~~~
\boldsymbol{\delta} v_m^{a} =
i \boldsymbol{\eta}_j \sigma_m \bar{\boldsymbol{\lambda}}^{ja}   
-i \boldsymbol{\lambda}_j^{\ a}{\sigma}_m\bar{\boldsymbol{\eta}}^j
\nonumber \\
\boldsymbol{\delta \lambda}_j^{\ a} &=& 
(\sigma^{mn} \boldsymbol{\eta}_j)v_{mn}^{a}+\sqrt{2}i(\sigma^m \bar{\boldsymbol{\eta}}_j) 
\mathcal{D}_m A^a+i(\boldsymbol{\tau} \cdot \boldsymbol{D}^a)_j^{\ k} \boldsymbol{\eta}_k
-\frac{1}{2} \boldsymbol{\eta}_j f^a_{\ bc} A^{*b} A^c,
\label{delta lambda}
\end{eqnarray}
where $\boldsymbol{\lambda}_i{}^a=({\lambda^a \atop \psi^a})$
and 
$\boldsymbol{\lambda}^{ia}=\epsilon^{ij}\boldsymbol{\lambda}_j{}^{a}$
(and similarly for the supersymmetry parameters $\eta_i$).
The $\boldsymbol{D}^a$ are auxiliary fields
\begin{eqnarray}
\boldsymbol{D}^a &=&\hat{\boldsymbol{D}}{}^a -\sqrt{2} g^{ab^*} 
\partial_{b^*}
\left( \boldsymbol{\mathcal{E}}A^{*0}+\boldsymbol{\mathcal{M}}
{\mathcal{F}}_0^* \right)\\
\boldsymbol{\mathcal{E}}&=&(0,\ -e,\ \xi),~~~
\boldsymbol{\mathcal{M}}=(0,\ -m,\ 0), \label{E and M}
\end{eqnarray}
where $\hat{\boldsymbol{D}}{}^a$
is the fermion bilinear terms of auxiliary fields
and is a real triplet under $SU(2)$.
Thus,
the $\CN=2$ transformation (\ref{delta lambda}) is 
$SU(2)$ covariant provided the two three-dimensional real vectors $\boldsymbol{\mathcal{E}}$ and $\boldsymbol{\mathcal{M}}$ 
transform as triplets. 
Their actual form (\ref{E and M}) tells us that
the rigid $SU(2)$ has been gauge fixed in this six-dimensional parameter space of ($\boldsymbol{\mathcal{E}}$, $\boldsymbol{\mathcal{M}}$), 
by making these two vectors point to a specific direction. 
The manifest $SU(2)$ covariance is lost at this point.

A very important property of the triplet of the auxiliary fields
$\boldsymbol{D}^a$ is that it is complex as opposed to be real.
Indeed, it has a constant imaginary part:
\begin{eqnarray}
\textrm{Im } \boldsymbol{D}^a = \delta^a_{\ 0} (-\sqrt{2} m)
\left(
0,
1,
0
\right). 
\end{eqnarray}
This supplies an essential ingredient for partial breaking of 
$\CN=2$ supersymmetry.

\section{Some Properties of the vacuum}
Because $\CF_a$ transforms in the adjoint representation
of $U(N)$, we fix $\CF$ of the form
\begin{eqnarray}
\CF=f(A^0)
+cA^0~\CG(\hat{B}) 
+\hat\CF(\hat A),
~~~
\hat A=A^{\hat a}t_{\hat a},
~~~
\hat{B}=\Tr(\hat A^2)/2c_2
\label{prepotential:solution}
\end{eqnarray}
where the indices $\hat a$ are for $SU(N)$
and
the constant $c_2$
is the quadratic Casimir.
Note that the $U(1)$ part and the $SU(N)$ part have non-trivial mixings
as long as $c\neq 0$.

Examining the scalar potential
\begin{eqnarray}
\CV
&=& g^{ab}\left(
\frac{1}{8}\fD_a\fD_b
+\xi^2\delta_{a}^0\delta_{b}^0
+\partial_a W\partial_{b^*} W^*
\right), \label{scalarpotential}
\end{eqnarray}
we find a stable minimum at $A^{\hat a}=0$ and
\begin{eqnarray}
f_{00}=
 -\frac{e}{m}
\pm i\frac{\xi}{m}
\label{minimum}
\end{eqnarray}
which represents the unbroken $SU(N)$  phase.
At this point,
the $U(1)$ fermion $\frac{1}{\sqrt{2}}(\lambda^0\mp\psi^0)$
and the $SU(N)$ fermions
$\frac{1}{\sqrt{2}}(\lambda^{\hat a}\mp\psi^{\hat a})$
remain massless,
while  the $U(1)$ fermion $\frac{1}{\sqrt{2}}(\lambda^0\pm\psi^0)$
and the $SU(N)$ fermions 
$\frac{1}{\sqrt{2}}(\lambda^{\hat a}\pm\psi^{\hat a})$
become massive with masses, 
$\left| -m\langle f_{000} \rangle \right|$
and $\left| -mc \langle \CG' \rangle\right|$, respectively.
Here, $\langle \cdots \rangle$ is the expectation value of
 $\cdots$ at the vacuum.
Because
\begin{eqnarray}
\left\langle
 \frac{\delta(\lambda^0 \mp \psi^0)}{\sqrt{2}}
\right\rangle = \mp 2mi(\eta_1 \pm \eta_2)
~,~~~~
\left\langle
 \frac{\delta(\lambda^0 \pm \psi^0)}{\sqrt{2}}
\right\rangle = 0,
\end{eqnarray}
the $U(1)$ massless fermion is regarded as the 
Nambu-Goldstone fermion.

In our model the $U(1)$ fermionic shift noted in ref \cite{CDSW} is realized on the vacuum as an approximate symmetry
coming from spontaneously broken supersymmetry.
\section{$\CN=2$ Supercurrents}
The conserved $R$ current $J$
 associated with the 
$U(1)_{R}$ transformation can be
 constructed when $\CF$ is
a degree two polynomial.
Though the $R$ current is not conserved with generic $\CF$,
we can construct the conserved $\CN=2$ supercurrents
of our model
as a broken $\CN=2$ supermultiplet of currents \cite{FZ}
 using the $R$ current.
The local $U(1)_R$ variation of $\CL$ implies that
\begin{eqnarray}
\partial_m 
\left
( -
\frac{1}{2}
\tr \bar{\sigma}^m J
\right
)
=i \left
( \sum_{n,j} (n-2) \frac{\partial}{\partial h_j^{(n)}} 
\right
) \mathcal{L} \equiv \Delta_h \mathcal{L},
\label{R variation}
\end{eqnarray}
where
 we write
  $\CF$ as $\sum_{n,j} h_j^{(n)}C_j^{(n)}(A^a)$ with 
$C_j^{(n)}(A^a)$ be $n$-th order $U(N)$ invariant polynomials 
in $A^a$,
and $h_j^{(n)}$ their coefficients. 
Acting the supersymmetry transformation on (\ref{R variation}),
and noting that $\boldsymbol{\delta}\CL=\partial_m X^m$ with
some $X^m$,
we obtain  a general construction of the conserved 
$\mathcal{N}=2$ supercurrents of our model;
\begin{eqnarray}
\boldsymbol{\eta}_j \boldsymbol{\mathcal{S}}^{(j)m}+\bar{\boldsymbol{\eta}}^j \bar{\boldsymbol{\mathcal{S}}}_{(j)}^{\ m} \equiv 
-{\frac{1}{2}} \tr (\bar{\sigma}^m \boldsymbol{\delta}J)-\Delta_h X^m . 
\label{supercurrent}
\end{eqnarray}
The form of the supercurrents given above tells us 
that our model does not permit a universal coupling to
 $\mathcal{N}=2$
supergravity. 
The piece $-\Delta_h X^m$ is not generic 
and depends on the functional form of the prepotential
$\mathcal{F}(A)$ in $A$. 
This
supports the point of view  
that $\mathcal{N}=2$ supersymmetric gauge models 
with nontrivial K\"ahler potential should be viewed 
as a low energy effective action.
Further acting the supersymmetry transformation on 
(\ref{supercurrent}), we can read off the constant matrix
$C^i_J$ in (\ref{intro})
as $C_i^{\ j}=+2m\xi (\boldsymbol{\tau}_1)_i^{\ j}$.

\section{General Analysis of the Vacua with Partial Supersymmetry Breaking}

Let us begin a more systematic analysis of the vacua with partial
supersymmetry breaking which applies not only the unbroken but also
the broken phase of $SU(N)$ gauge symmetry.
Let us convert the scalar potential (\ref{scalarpotential}) into
\begin{eqnarray}
\mathcal{V}=\frac{1}{8} g_{bc} \mathfrak{D}^b \mathfrak{D}^c 
+ \frac{1}{2} g^{bc} \tilde{\boldsymbol{D}}_b^* \cdot \tilde{\boldsymbol{D}}_c \ \ ,
\end{eqnarray}
where
\begin{eqnarray}
\tilde{\boldsymbol{D}}_b = \sqrt{2} \left( 0,\partial_{b^*} W^*,-\xi \delta_b^{\ 0} \right)
\label{D tilde}
\end{eqnarray}
is the bosonic contribution of the auxiliary field $\boldsymbol{D}^a$ with its index lowered by $g_{ba}$.
Noting that
$\left< \mathfrak{D}^a \right>=0$,
which follows from $\left< A^r \right>=0$,
where the index $r$ is for
non-Cartan generators of $u(N)$,
we derive the following condition on the vacuum;
\begin{eqnarray}
0&=&-4i \left< \partial \mathcal{V}/\partial A^a \right> =\left< \mathcal{F}_{abc} \tilde{\boldsymbol{D}}^b \cdot \tilde{\boldsymbol{D}}^c \right> 
\nonumber \\
 &=&2 \left< \mathcal{F}_{abc} g^{bb'} g^{cc'} (\Re \partial_{b'} W)(\Re \partial_{c'} W) \right> 
\nonumber \\ & &
+4im \left< \mathcal{F}_{ab0}g^{bb'} (\Re \partial_{b'} W) \right>
 +2 \left< \mathcal{F}_{abc} g^{b0} g^{c0} \right> \xi^2 -2 \left< \mathcal{F}_{a00} \right> m^2. \label{four}
\end{eqnarray}
This includes the case of the vacuum with unbroken $SU(N)$ gauge symmetry as that satisfying
$\left< \partial_a W \right>=\delta_a^{~ 0} (e+m \left< f_{00} \right>)$, and $\big< g^{\hat{b} 0} \big>=0$.

Let us now turn to analyze the condition of partial breaking of
 $\mathcal{N}=2$ supersymmetry. We have
\begin{eqnarray}
\left< \boldsymbol{\delta} \boldsymbol{\lambda}_j^{\ a} \right>=i \left< ( \boldsymbol{\tau} \cdot \tilde{\boldsymbol{D}}^a )_j^{\ k} \boldsymbol{\eta}_k \right>.
\end{eqnarray}
In order to find a Nambu-Goldstone fermion from the $\mathcal{N}=2$ doublets of $U(N)$ fermions, we must find a set of nonvanishing coefficients $C_a$ such that
\begin{eqnarray}
&&
\tilde{\boldsymbol{D}}(C_a) \equiv \sum_a C_a \tilde{\boldsymbol{D}}^a ,
\nonumber\\
&&
0= \left< \det \boldsymbol{\tau} \cdot \tilde{\boldsymbol{D}} \right>=\left< \tilde{\boldsymbol{D}} \cdot \tilde{\boldsymbol{D}} \right>. \label{seven}
\end{eqnarray}
It is straightforward to spell out (\ref{seven}) in terms of the real and imaginary parts of $\tilde{\boldsymbol{D}}_a$,
using (\ref{D tilde}).

Let us further assume
\begin{eqnarray}
\left< \Re \partial_b W \right>=0~.
\label{eight}
\end{eqnarray}
Eq.(\ref{four}) then gives us
 (for non-zero $\left< \mathcal{F}_{a00} \right>$)
\begin{eqnarray}
\frac{\left< \mathcal{F}_{abc}\right> \left< g^{b0} \right> \left< g^{c0} \right>}{\left< \mathcal{F}_{a00} \right>}
=\frac{m^2}{\xi^2}~~~~~~(\mbox{no sum on}~~ a).
 \label{nine}
\end{eqnarray}
On the other hand, the condition (\ref{seven}) gives us
\begin{eqnarray}
\left( \sum_b \frac{C_b}{C_0} \left< g^{b0} \right> \right)^2=\frac{m^2}{\xi^2}~.
 \label{ten}
\end{eqnarray}
Eq.(\ref{eight}),(\ref{nine}) and the existence of the coefficients $C_b$ satisfying (\ref{ten}) are the conditions that 
the vacuum state satisfies and include the vacuum discussed in \S 4.


\section{Acknowledgements}
This work is supported in part by the Grant-in-Aid for Scientific
Research(16540262) from the Ministry of Education,
Science and Culture, Japan,
and by
the 21 century COE program
``Constitution of wide-angle mathematical basis focused on knots".

\bibliographystyle{plain}

\end{document}